\documentclass[twoside,amssymb,aps,showpacs,showkeys]{revtex4}

\usepackage{dcolumn}

\usepackage{graphicx} 

\usepackage{bm}

\begin{document}

\title{Karhunen-Lo\`eve Basis Functions of Kolmogorov Turbulence in the Sphere}

\author{Richard J. Mathar} 
\homepage{http://www.strw.leidenuniv.nl/~mathar} 
\email{mathar@strw.leidenuniv.nl} 
\affiliation{Leiden Observatory, Leiden University, P.O. Box 9513, 2300 RA Leiden, The Netherlands} 


\date{\today} 

\begin{abstract} 
The
statistically independent Karhunen-Lo\`eve modes of refractive indices with
isotropic Kolmogorov spectrum of the covariance are calculated inside a sphere of given
radius, rendered as series
of 3D Zernike functions. Many of the symmetry arguments of the associated
2D problem for the circular input pupil remain valid.
The technique of efficient diagonalization of the eigenvalue
problem in wavenumber space is founded on the
Fourier representation of the 3D Zernike basis.
\end{abstract}

\maketitle  
\section{Overview}

\subsection{Model of the Refractive Index Structure Function} \label{sec.intro}
This paper describes modeling of variations of refractive indices $n$
based on their
structure functions $\mathcal{D}$
\begin{equation}
\mathcal{D}_n(\Delta\mathbf{r})=C_n^2 |\Delta \mathbf{r}|^\gamma
\end{equation}
with Kolmogorov power index
\begin{equation}
\gamma=2/3,
\end{equation}
sampled at points which are $\Delta \mathbf{r}$ apart and located in an air volume
between the light source and some input pupil of a receiver.
The familiar assumptions of this model are
\begin{itemize}
\item
``single layer''  homogeneity: the refractive index structure constant $C_n^2$
is indeed constant
over the entire volume that is probed by the electromagnetic wave,
\item
isotropy: the structure function depends on the absolute value of the
vector $\Delta \mathbf{r}$, but not on its direction.
\end{itemize}
In the conjugated Fourier variable $\bm{f}$ of wave numbers, the turbulent
spectrum is represented by the
(again isotropic) Wiener spectrum $\Phi_n(f)$
\cite{NollJOSA66,PerezOptLett33}.
The expectation value of the mean
appears as a constant in the Fourier
kernel, and the expectation value of the covariance
as a cosine:
\begin{equation}
\mathcal{D}_n(\Delta \mathbf{r})=2\int
\Phi_n(f)[1-\cos(2\pi\bm{f}\cdot \Delta \mathbf{r})]
d^3f
.
\end{equation}
[The
imaginary part of the factor $\exp(2\pi i \bm{f}\cdot \Delta \bm{r})$,
which is induced by the shift theorem Fourier transforms and which represents
the contributions with odd parity, vanishes by isotropy after integration.]
The power spectrum is
\cite{ConanJOSAA25,Conan00}
\begin{equation}
\Phi_n(f)
=
-
\pi^{3/2-\beta}
\frac{ \Gamma(\beta/2) }{ 2\Gamma(\frac{3-\beta}{2}) }
C_n^2 f^{-\beta}
 \equiv c_\Phi C_n^2 f^{-11/3}
;\quad
c_\Phi  \approx 0.00969315 
,
\label{eq.Phi3D}
\end{equation}
at a power spectral index
\begin{equation}
\beta = \gamma+3\quad (=11/3).
\end{equation}
Where the variable $\bm{\kappa}=2\pi\bm{f}$ is used in lieu of
$\bm{f}$, the constant $(2\pi)^\gamma c_\Phi\approx 0.033005$ takes the role of $c_\Phi$.

\subsection{Circular (Projected) 2D Case}

Above, the structure function and its spectrum are defined as functions
of the 3D distances $\Delta r$ and $\sigma$.
Integration along two parallel  lines of sight
through a layer of vertical thickness $h$ at an air mass $1/\sin a$
at a projected baseline $P$ for momentum number $k$ (equal to $2\pi$ divided
through the wavelength of the electromagnetic wave)
defines 2D phase structure functions
\begin{equation}
\mathcal{D}_\varphi(P)=C_n^2 k^2 g(\gamma) P^{\gamma+1}\frac{h}{\sin a}
,
\end{equation}
where
\begin{equation}
g(\gamma)=\sqrt{\pi}\frac{\Gamma(-(1+\gamma)/2)}{\Gamma(-\gamma/2)}
\approx 2.9144,\quad (\gamma=2/3)
\label{eq.consth}
\end{equation}
is a geometric constant from the integration
\cite{RoddierProgOpt19,FriedJOSA55,HufnagelJOSA54,StrohbeinProgOpt9}.
The information sampled along the pencils of length $h/\sin a$
through the layer  is reduced to $\mathcal{D}_\varphi$, a function of the two 
components of $P$ in the input pupil of the receiver.
Noll's notation $\Phi_\varphi(\nu)=0.023r_0^{-5/3} \nu^{-11/3}$
of the associated 2D power spectrum of the phase differences
in terms of the Fried parameter $r_0$---at the wavelength $2\pi/k$ of the 
electromagnetic spectrum---
\cite[(21)]{NollJOSA66}
is equivalent as we note that
$0.023$ is the product of the ad-hoc scaling parameter
\cite{FriedJOSA56_1372}
\begin{equation}
2[(24/5)\Gamma(6/5)]^{5/6}\approx 6.883877
\end{equation}
by
\begin{equation}
-\frac{\Gamma(\beta/2)}{
2\pi^{2+\gamma}
\Gamma(1-\beta/2)
}\approx 0.0033260
.
\label{eq.constphi2d}
\end{equation}
The association of the constants in equations (\ref{eq.Phi3D}), (\ref{eq.consth})
and (\ref{eq.constphi2d}) is summarized in the following diagram. The role
of $g(\gamma)$ switches from multiplier to divisor while transiting
from real to wavenumber space.
\begin{center}
\setlength{\unitlength}{1cm}
\begin{picture}(8.5,4.0)
\put(1.0,3.5){$C_n^2$}
\put(2.5,3.5){\vector(1,0){5.2}}
\put(3.5,3.0){3D Fourier Transform}
\put(0.5,2.5){integration over $h$}
\put(0.5,2.0){multiply $g(\gamma)$}
\put(1.0,3.3){\vector(0,-1){2.2}}
\put(1.0,0.5){$2.914C_n^2$}
\put(2.5,0.5){\vector(1,0){5.2}}
\put(3.5,0.8){2D Fourier Transform}
\put(8.0,3.5){$0.00969C_n^2$}
\put(8.0,3.3){\vector(0,-1){2.2}}
\put(8.0,0.5){$0.00333C_n^2$}
\put(7.5,2.0){divide $g(\gamma)$}
\end{picture}
\end{center}
The intend is to show how the phase structure function is associated
with integration of a refractive index structure function over path lengths; the Fried
parameter and some of the associated geometric constants will disappear from the
analysis of the next section, where no such path integration is performed.

This concludes the overview of the literature. Karhunen-Lo\`eve (KL)
modes in the projected (2D) case are not reviewed
here; we refer to the available literature
\cite{FriedJOSA68,DaiJOSAA12,RoddierOE29,MatharArxiv0705b}.

\subsection{Applications}
The aim of calculations in section \ref{sec.main} is to remove the constraint of
working with optical path lengths integrated along two thin, parallel
rays through the turbulent atmosphere.
Simulations of anisoplanatism, multi-conjugate adaptive optics, and atmospheric
transfer functions for radio (as opposed to optical) receivers often integrate
over extended fields-of-view.
Essentially we factorize the problem;
we provide means to fill a volume with instances of a turbulent refractive index,
but keep aside
by which weighing scheme sub-volumes accumulate the instrumental signal.

To keep mathematics simple, this volume is given spherical shape, the shape
that is symmetry-adapted to the isotropy of the structure function. As one
extreme application of the geometry one could image the center of this sphere
at the center of the Earth, the sphere radius set to the Earth radius plus
some tenth of kilometers, and
then evaluating refractive indices only in the thin inner hull close to the
sphere surface---representing the entire Earth's
atmosphere---computed
from superposition of the KL modes.

\section{Karhunen-Lo\`eve Functions in the Sphere} \label{sec.main}

\subsection{Spherical Basis} \label{sec.spherFT}

The generic form of the Karhunen-Lo\`eve eigenvalue equation
for the modes $K$ and their eigenvalues ${\cal B}^2$
associated with a covariance $C$ sampled over a
spherical region of radius $R$ is
\begin{equation}
\int_{r'\le R} C(|\mathbf{r}-\mathbf{r}'|) K(\bm{r'}) d^3r'= {\cal B}^2  K(\bm{r}).
\end{equation}
One benefit earned from the constraints of the model described in
section \ref{sec.intro} is that a radial scaling
\begin{equation}
r=Rx
\end{equation}
maps these cases onto a single unified model inside the unit sphere:
\begin{equation}
\int_{x'\le 1} |\mathbf{x}-\mathbf{x}'|^\gamma K(\mathbf{x'}) d^3x'=\frac{{\cal B}^2}{C_n^2 R^{\gamma+3}}  K(\mathbf{x}).
\end{equation}
The other benefit is that the spherical symmetry of the domain of integration matches
the isotropy in the covariance; moving on to the Fourier domain of the dimensionless
$\bm{\sigma}=R\bm{f}$,
\begin{equation}
K(x')=\int d^3\sigma K(\bm{\sigma}) e^{-2\pi i \bm{\sigma}\cdot \mathbf{x'}}
\end{equation}
and
\begin{equation}
C(|x-x'|)=\int d^3\sigma C(\sigma) e^{-2\pi i \bm{\sigma}\cdot (\mathbf{x}-\mathbf{x}')}
\end{equation}
transforms the convolution into a product, in which the eigenvalue problem
is block diagonal in the angular variables
\cite{RoddierOE29},
\begin{equation}
C(\sigma)
K({\bf \sigma})
=\frac{{\cal B}^2}{R^{\gamma+3}}  K(\sigma)
.
\label{eq.KLsigma}
\end{equation}
With the same argument as for the 2D case, ie, because the exponent
of $C\propto \sigma^{-\beta}$ is too large for the Kolmogorov spectrum
to remain integrable in the large-scale limit $\sigma\to 0$, the decomposition (\ref{eq.KLsigma})
is not possible for functions with nonzero $K(\sigma)$ at the origin of the $\sigma$-space:
the
homogeneous mode, the analog to the 2D piston mode, is left out of the analysis.
${\cal D}_n$ does not convey information on the mean of the refractive index.

\subsection{Diagonalization in Wavenumber Space}

Subject to the usual ambiguity of any choice of basis functions, we expand
the $K(x)$ in an orthogonal Zernike basis, suggested in Appendix \ref{app.Z},
serializing the eigenfunctions for each block $l$ of the eigenvalue problem by an index $p$:
\begin{equation}
K_p^{(l)}(\bm{\sigma})=\sum_n \kappa_{n,l,p}Z_{nl}^{(m)}(\bm{\sigma})
=
4\pi \sum_n \kappa_{n,l,p}i^l Y_l^m(\theta_\sigma,\phi_\sigma) R_n^{(l)}(\sigma)
.
\label{eq.KofZ}
\end{equation}
The reduction of (\ref{eq.KLsigma}) with (\ref{eq.Phi3D}) reads
\begin{equation}
c_\Phi \sigma^{-\beta}K_p^{(l)}(\sigma)=\lambda_{p,l}^2 K_p^{(l)}(\sigma);
\quad \lambda_{p,l}^2= \frac{{\cal B}^2}{R^\beta C_n^2}.
\label{eq.lsq}
\end{equation}
Formulation of an eigenvalue problem of matrix algebra proceeds
with (\ref{eq.Rsigortho}),
\begin{equation}
c_\Phi
\sum_{n} \kappa_{n,l,p}
\int_0^\infty \sigma^2 \sigma^{-\beta}
R_n^{(l)}(\sigma)
R_{n'}^{(l)}(\sigma)
d\sigma
=
  \frac{\lambda_{p,l}^2}{(4\pi)^2} \kappa_{n',l,p}
\end{equation}
A core integral matrix $I_{n,n'}^{(l)}$ is defined to tighten the notation,
\begin{equation}
c_\Phi
\int_0^\infty \sigma^2 \sigma^{-\beta}
R_n^{(l)}(\sigma)
R_{n'}^{(l)}(\sigma)
d\sigma
\equiv I_{nn'}^{(l)}
  \frac{1}{(4\pi)^2}
.
\label{eq.Idef}
\end{equation}
The integral over the product of
Bessel Functions of the form (\ref{eq.Rnlsigma}) yields
\cite[6.574.2]{GR}\cite[p. 49]{MO2Afl}
\begin{equation}
I_{nn'}^{(l)}
=
c_\Phi
 \frac{\pi^\beta}{2} \sqrt{(2n+3)(2n'+3)}
(-1)^{(n-n')/2}
\frac{
\Gamma(\beta+1)
\Gamma(\frac{n+n'+3-\beta}{2})
}{
\Gamma(\frac{n-n'+\beta}{2}+1)
\Gamma(\frac{n'+n+\beta+5}{2})
\Gamma(\frac{n'-n+\beta}{2}+1)
}
\label{eq.Inn}
\end{equation}
for $n,n'=l,l+2,l+4,\ldots$.
These are the matrices to be diagonalized to establish the KL basis functions,
one matrix for each $l$,
\begin{equation}
\sum_n I_{n,n'}^{(l)} \kappa_{n,l,p} =\lambda_{p,l}^2 \kappa_{n',l,p}.
\end{equation}
The upper right corners of the symmetric $I_{n,n'}^{(l)}=I_{n',n}^{(l)}$
are shown in Table \ref{tab.Inn0} for even and in Table \ref{tab.Inn1} for odd $l$.
The individual array for a specific $l$ is the sub-array in which all rows
$n <l$ and columns $n'<l$ are removed.
A common factor has been split off to write them as exact fractions.
If no
such scaling is chosen, tables like Table \ref{tab.Inn1too} result.
As argued at the end of
section \ref{sec.spherFT},
row $n=0$ and column $n'=0$
in Table \ref{tab.Inn0} are deleted prior to the diagonalization.

\begin{table}
\caption{
$I_{nn'}^{(l)} /[2^{2/3}\pi \sqrt{(2n+3)(2n'+3)}]$
for $n,n'\le 12$, $l$ even.
Multiply by $\sqrt{(2n+3)(2n'+3)}$ and in addition 
by $2^{2/3}\pi \approx 4.986967$ to obtain $I_{nn'}^{(l)}$.
The table is symmetric; only the non-redundant upper right triangle is shown.
}
\label{tab.Inn0}
\begin{ruledtabular}
\begin{tabular}{l|ccccccc}
    & 0 & 2 & 4 & 6 & 8 & 10 & 12
\\ \hline
0  & $-{\frac {54}{385}}
$ & $-{\frac {54}{7735}}
$ & ${\frac {27}{142324}}
$ & ${\frac {135}{78420524}}
$ & ${\frac {27}{215656441}}
$ & ${\frac {27}{1545789175}}
$ & ${\frac {27}{7647588550}}
$\\2  &  & ${\frac {27}{20020}}
$ & $-{\frac {27}{117572}}
$ & ${\frac {135}{7436429}}
$ & ${\frac {27}{98025655}}
$ & ${\frac {27}{980256550}}
$ & ${\frac {27}{5637584050}}
$\\4  &  &  & ${\frac {27}{209209}}
$ & $-{\frac {27}{734825}}
$ & ${\frac {27}{6760390}}
$ & ${\frac {27}{357505330}}
$ & ${\frac {27}{3038795305}}
$\\6  &  &  &  & ${\frac {27}{950950}}
$ & $-{\frac {27}{2679950}}
$ & ${\frac {27}{20957209}}
$ & ${\frac {27}{977699359}}
$\\8  &  &  &  &  & ${\frac {27}{2947945}}
$ & $-{\frac {621}{168568855}}
$ & ${\frac {27}{51867340}}
$\\10  &  &  &  &  &  & ${\frac {621}{167806100}}
$ & $-{\frac {18009}{11151478100}}
$\\12  &  &  &  &  &  &  & ${\frac {3132}{1803915575}}
$\\
\end{tabular}
\end{ruledtabular}
\end{table}
\begin{table}
\caption{
$I_{nn'}^{(l)} /[2^{2/3}\pi\sqrt{(2n+3)(2n'+3)}]$
for $n,n'\le 13$, $l$ odd.
}
\label{tab.Inn1}
\begin{ruledtabular}
\begin{tabular}{l|ccccccc}
    & 1 & 3 & 5 & 7 & 9 & 11 & 13
\\ \hline
1  & ${\frac {54}{5005}}
$ & $-{\frac {27}{30940}}
$ & ${\frac {135}{2704156}}
$ & ${\frac {135}{215656441}}
$ & ${\frac {27}{490128275}}
$ & ${\frac {27}{3091578350}}
$ & ${\frac {27}{13945602650}}
$\\3  &  & ${\frac {27}{76076}}
$ & $-{\frac {27}{323323}}
$ & ${\frac {27}{3380195}}
$ & ${\frac {27}{196051310}}
$ & ${\frac {27}{1787526650}}
$ & ${\frac {27}{9583892885}}
$\\5  &  &  & ${\frac {27}{475475}}
$ & $-{\frac {27}{1469650}}
$ & ${\frac {27}{12327770}}
$ & ${\frac {27}{607759061}}
$ & ${\frac {27}{4888496795}}
$\\7  &  &  &  & ${\frac {459}{29479450}}
$ & $-{\frac {27}{4555915}}
$ & ${\frac {27}{33713771}}
$ & ${\frac {27}{1504152860}}
$\\9  &  &  &  &  & ${\frac {621}{109073965}}
$ & $-{\frac {621}{259336700}}
$ & ${\frac {783}{2230295620}}
$\\11  &  &  &  &  &  & ${\frac {18009}{7215662300}}
$ & $-{\frac {3132}{2787869525}}
$\\13  &  &  &  &  &  &  & ${\frac {3132}{2525481805}}
$\\
\end{tabular}
\end{ruledtabular}
\end{table}
\begin{table}
\caption{
$I_{nn'}^{(1)}$ limited to $n,n'\le 11$.
This repeats the values of Table \ref{tab.Inn1} without the scale factor.
}
\label{tab.Inn1too}
\begin{ruledtabular}
\begin{tabular}{l|dddddd}
    & 1 & 3 & 5 & 7 & 9 & 11
\\ \hline
1  & $0.269027$ & $-0.029194$ & $0.002007$ & $0.000029$ & $0.000003$ & $0.000000$\\
3  &  & $0.015929$ & $-0.004505$ & $0.000493$ & $0.000009$ & $0.000001$\\
5  &  &  & $0.003681$ & $-0.001362$ & $0.000180$ & $0.000004$\\
7  &  &  &  & $0.001320$ & $-0.000558$ & $0.000082$\\
9  &  &  &  &  & $0.000596$ & $-0.000274$\\
11  &  &  &  &  &  & $0.000311$\\
\end{tabular}
\end{ruledtabular}
\end{table}

As already in the 2D case, the relative strength of values on the diagonals
compared to the sub-diagonals is predictive of how close the KL modes
resemble pure Zernike functions.
An explicit table of the dominant modes is given in Appendix \ref{sec.modes},
sorted with respect to decreasing $\lambda_{p,l}^2$.

\subsection{Real-valued Basis}

In practice, the component $e^{im\phi}$ of the spherical harmonics in
the Zernike basis (\ref{eq.Ylm}) may be replaced by switching in the real-valued
functions
\cite{Bradley1972}
\begin{equation}
\sqrt{\epsilon_m}\cos(m\theta),\quad \sqrt{\epsilon_m}\sin(m\theta),
\end{equation}
where 
\begin{equation}
\epsilon_m \equiv \left\{
\begin{array}{ll} 1, & m=0 \\ 2, & m\ge 1\end{array}
\right.
\end{equation}
is the Neumann factor. This choice maintains normalizations, and maintains
the $2l+1$ fold degeneracy of the Karhunen-Lo\`eve modes.

\section{Summary} 
This work traced the well-known Karhunen-Lo\`eve functions of
a covariance with power spectral index 5/3 over a circular telescope entrance
pupil back to the associated functions with power spectral index 2/3 over the sphere.
The solution of the eigenvalue integral equation for the covariance has
been formulated for expansions in a 3D Zernike basis.
Bare Zernike functions turn out to be
good approximations to the large scale modes.

\begin{acknowledgments}
This work is supported by the NWO VICI grant
639.043.201
to A. Quirrenbach,
``Optical Interferometry: A new Method for Studies of Extrasolar Planets.''
\end{acknowledgments}

\appendix

\section{3D Zernike Basis}\label{app.Z}

\subsection{Definition}

We define an orthogonal 3D Zernike Basis $Z_{n,l}^{(m)}(r,\theta,\phi)$
over the unit sphere in strict analogy to the domain
of the unit circle
\cite{BhatiaPCRS50,BhatiaPRSB65}
as a product over (i) Courant-Hilbert Jacobi polynomials $G$ to span the radial
coordinate $x$ and (ii) Spherical Harmonics $Y_l^m$  to cover the
azimuthal and polar angles $\phi$ and $\theta$:
\begin{equation}
Z_{n,l}^{(m)} \equiv R_n^{(l)}(x) Y_l^m(\theta,\phi),
\quad
0\le x\le 1,\quad 0\le\theta\le\pi,\quad 0\le\phi\le 2\pi.
\end{equation}
Spherical Harmonics
\begin{equation}
Y_l^m = \sqrt{\frac{(2l+1)(l-|m|)!}{4\pi(l+|m|)!}} P_l^{|m|}(\cos\theta)e^{im\phi},
\quad -l\le m \le l,
\label{eq.Ylm}
\end{equation}
are a product of $e^{im\phi}\sin^{|m|}\theta$ with a polynomial in
$\cos\theta$ of degree $l-|m|$.
(There are at least three different sign/phase conventions in the literature,
of which only one is picked by this equation. Since the manuscript does not deal with
aspects of angular momentum coupling, the specific choice is not relevant.)
The transformation to Cartesian coordinates $(x\sin\theta\cos\phi,x\sin\theta\sin\phi,
x\cos\theta)$ has been discussed in the literature
\cite{MatharIJQC90,NovotniCAD36}.

\begin{figure}
\includegraphics[width=8cm,angle=0]{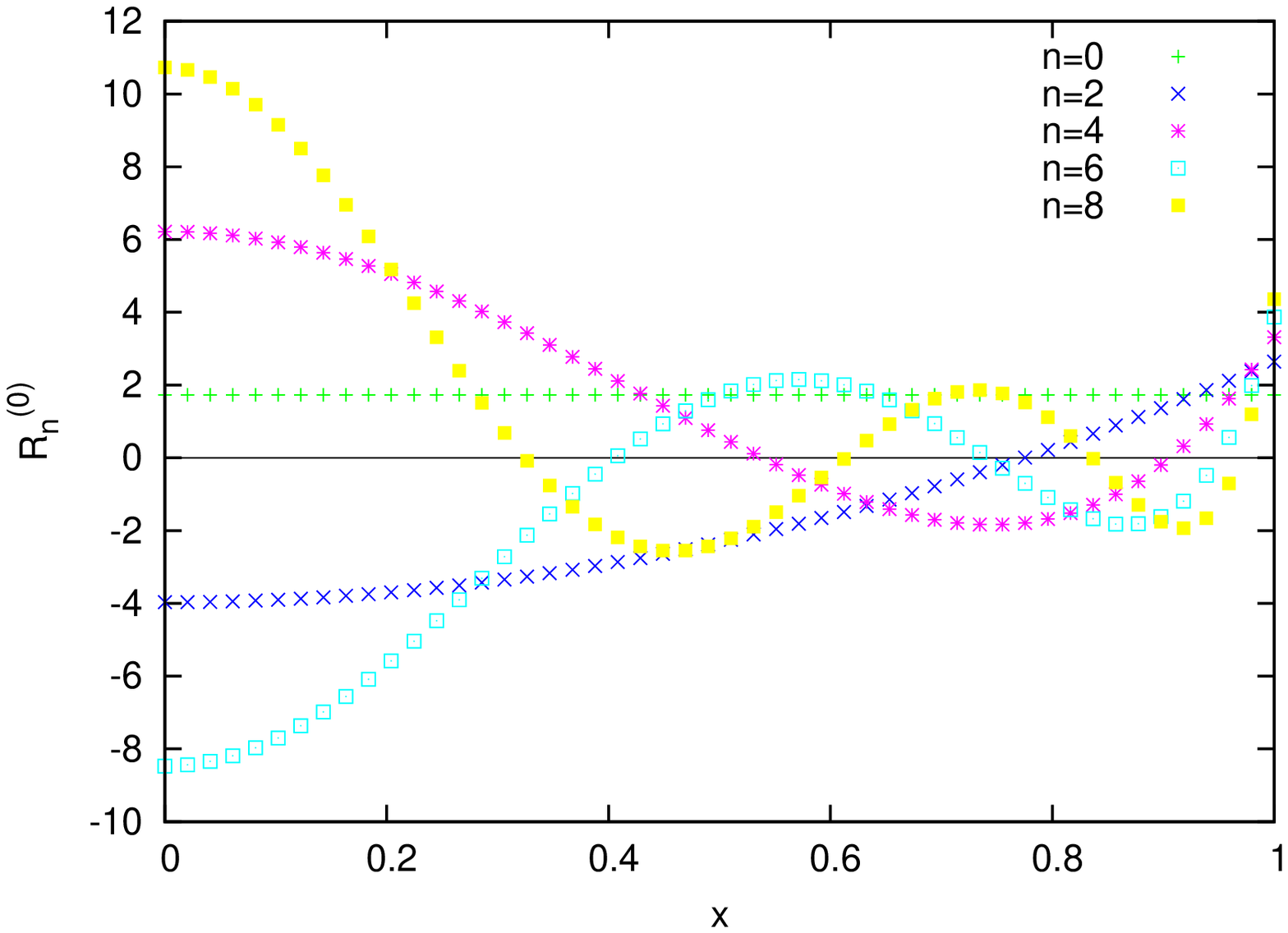}
\includegraphics[width=8cm,angle=0]{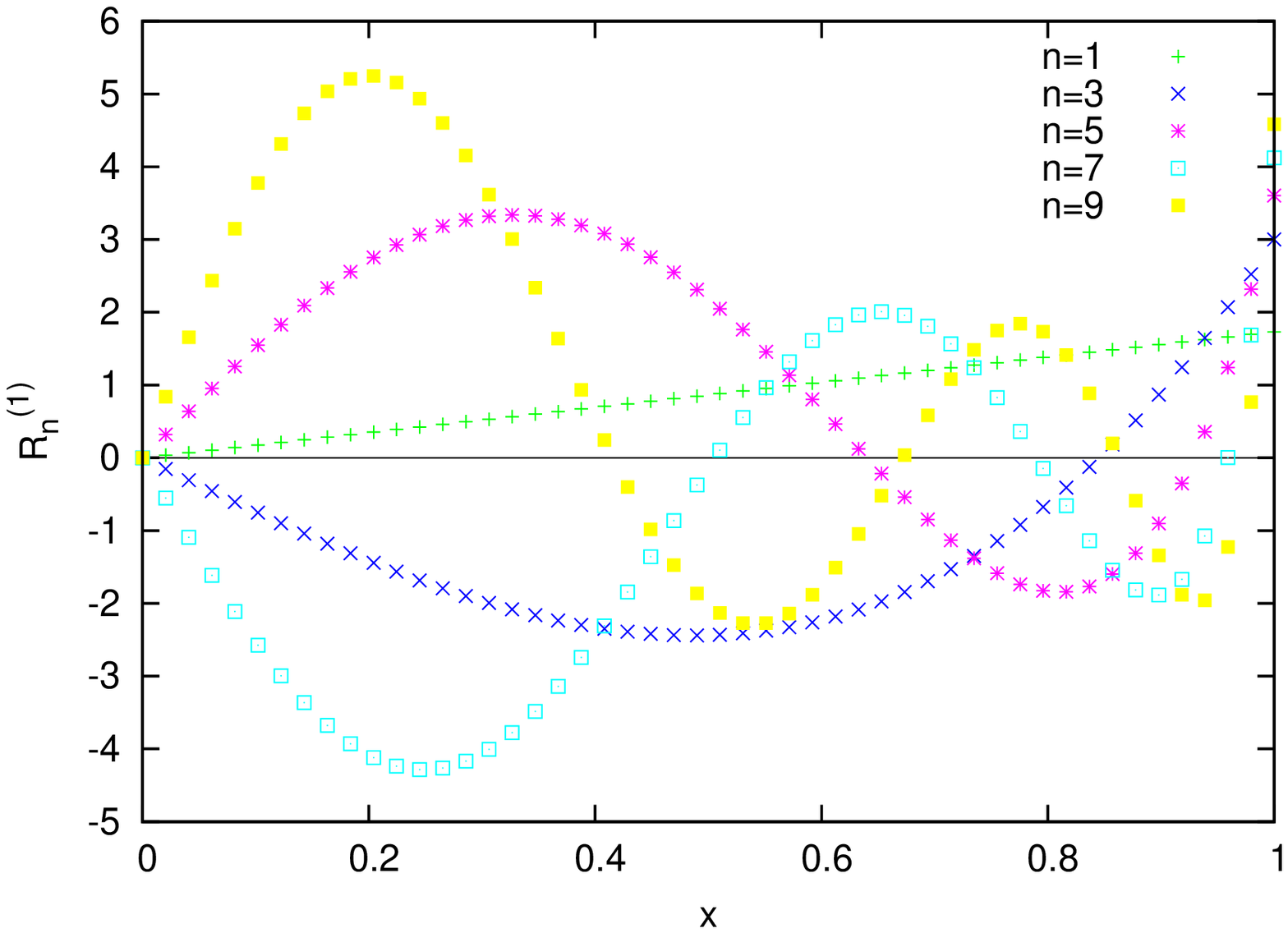}
\caption{Radial part $R_n^{(l)}(x)$
of the 3D Zernike basis defined in (\ref{eq.Rnorm}), $l=0$ or $1$.
}
\label{fig.R}
\end{figure}

Rotation of the 3D coordinate system mixes the components of $Y_l^m$,
keeping $l$ constant
\cite[2.1.3]{Bradley1972}\cite{AltmannPC53,Edmonds}.
The role of the azimuthal number $m$ of the 2D case
\cite{NollJOSA66}
is therefore taken over by $l$ to set the lowest order
to the radial polynomials
\cite{BhatiaPCRS50,NovotniACM,NovotniCAD36}.
With this constraint, demanding orthogonality with respect to 
the weight function $x^2$ in $0\le x\le 1$
(of the Jacobian of the spherical coordinates), and setting a smoothness-constraint
at the origin $x=0$ in the sense that the powers of $x$ in the polynomial
are all either even or odd, the radial functions need to be
proportional to $x^l G_\alpha(q,q,x^2)$,
where
\begin{eqnarray}
G_\alpha(q,q,y)
&=&
\frac{\Gamma(q+\alpha)}{\Gamma(q+2\alpha)}
\sum_{k=0}^\alpha (-1)^k{\alpha \choose k}\frac{\Gamma(q+2\alpha-k)}{\Gamma(q+\alpha-k)}y^{\alpha-k}
\\
&=&
\frac{\Gamma(q+\alpha)}{\Gamma(q+2\alpha)}
(-1)^\alpha \sum_{k=0}^\alpha \frac{\Gamma(\alpha+1)}{\Gamma(\alpha-k+1)k!}\frac{\Gamma(q+\alpha+k)}{\Gamma(q+k)}(-y)^k
\\
&=&
\frac{\Gamma^2(q+\alpha)}{\Gamma(q+2\alpha)\Gamma(q)}(-1)^\alpha \, _2F_1(-\alpha,q+\alpha;q;y)
=
(-1)^\alpha
\frac{ {q+\alpha-1 \choose \alpha} }{ {q+2\alpha-1 \choose \alpha} }
\, _2F_1(-\alpha,q+\alpha;q;y)
\end{eqnarray}
are Courant-Hilbert Jacobi polynomials
\cite[22.3.3]{AS}
with parameters
\begin{equation}
\alpha \equiv (n-l)/2,\quad
q \equiv l+3/2.
\end{equation}
The characteristic difference with the 2D
case, see \cite{MatharArxiv0705a} and references therein,
is that the parameter
$q$ has increased by $1/2$ to cope with the higher power of the
Jacobian of the Cartesian-to-spherical (as opposed to Cartesian-to-circular)
transformation of coordinates.
[This boost of the parameter remains visible in the matrix elements
(\ref{eq.Inn}) and remains visible in the Bessel Function indices (\ref{eq.Rnlsigma}) which
are translated from integer to half-integer values.]
Ortho-normalization based on their orthogonality
\cite[22.2.2]{AS}
yields
\begin{eqnarray}
R_n^{(l)}(x)
&\equiv&
\sqrt{2n+3}
\frac{
  \left(\frac{n+l}{2}+1\right)! (2n+2)! 
  }{
  2^{n-l}(n+l+2)!\left(\frac{n-l}{2}\right)!(n+1)!
  }
x^lG_{(n-l)/2}(l+\frac{3}{2},l+\frac{3}{2},x^2)
\label{eq.Rnorm}
\\
&=&
\sqrt{2n+3}\frac{\Gamma(2\alpha+q)}{\Gamma(\alpha+q)\Gamma(\alpha+1)}x^l G_\alpha(q,q,x^2)
,
\end{eqnarray}
\begin{equation}
\int_0^1 x^2 R_n^{(l)}(x) R_{n'}^{(l)}(x) dx = \delta_{n,n'}.
\end{equation}
Some examples of these polynomials are given in
Table \ref{tab.R} for comparison with the 2D case
\cite[Table I]{NollJOSA66}.
A subset is also visualized in Figure \ref{fig.R}
for comparison with \cite[Fig.\ 1]{MakJMGM26}.
The nodal counts are
as expected for any of the classical orthogonal polynomials.
The overall choice of signs,
ie, $R_n^{(l)}(1)>0$, is just inherited from the Jacobi polynomials, and carries no
physical significance.

\begin{table}
\caption{
The radial polynomials $R_n^{(l)}(x)$ for $l\le 4$ and $n\le 10$.
}
\label{tab.R}
\begin{ruledtabular}
\begin{tabular}{rrc}
$l$ & $n$ & $R_n^{(l)}(x)$ \\
\hline
0 & 0 & $\sqrt {3}
$ \\ 
0 & 2 & $\frac{1}{2}\sqrt {7} \left( -3+5\,{x}^{2} \right) 
$ \\ 
0 & 4 & $\frac{1}{8}\sqrt {11} \left( 15+63\,{x}^{4}-70\,{x}^{2} \right) 
$ \\ 
0 & 6 & $\frac{1}{16}\sqrt {15} \left( -35+429\,{x}^{6}-693\,{x}^{4}+315\,{x}^{2} \right) 
$ \\ 
0 & 8 & ${\frac {1}{128}}\,\sqrt {19} \left( 315+12155\,{x}^{8}-25740\,{x}^{6}+18018\,{x}^{4}-4620\,{x}^{2} \right) 
$ \\ 
0 & 10 & ${\frac {1}{256}}\,\sqrt {23} \left( -693+88179\,{x}^{10}-230945\,{x}^{8}+218790\,{x}^{6}-90090\,{x}^{4}+15015\,{x}^{2}
\right) 
$ \\ 
1 & 1 & $x\sqrt {5}
$ \\ 
1 & 3 & $\frac{3}{2} x \left( -5+7\,{x}^{2} \right) 
$ \\ 
1 & 5 & $\frac{1}{8}\sqrt {13}x \left( 35+99\,{x}^{4}-126\,{x}^{2} \right) 
$ \\ 
1 & 7 & $\frac{1}{16}\sqrt {17}x \left( -105+715\,{x}^{6}-1287\,{x}^{4}+693\,{x}^{2} \right) 
$ \\ 
1 & 9 & ${\frac {1}{128}}\,\sqrt {21}x \left( 1155+20995\,{x}^{8}-48620\,{x}^{6}+38610\,{x}^{4}-12012\,{x}^{2} \right) 
$ \\ 
2 & 2 & ${x}^{2}\sqrt {7}
$ \\ 
2 & 4 & $\frac{1}{2}\sqrt {11}{x}^{2} \left( -7+9\,{x}^{2} \right) 
$ \\ 
2 & 6 & $\frac{1}{8}\sqrt {15}{x}^{2} \left( 63+143\,{x}^{4}-198\,{x}^{2} \right) 
$ \\ 
2 & 8 & $\frac{1}{16}\sqrt {19}{x}^{2} \left( -231+1105\,{x}^{6}-2145\,{x}^{4}+1287\,{x}^{2} \right) 
$ \\ 
2 & 10 & ${\frac {1}{128}}\,\sqrt {23}{x}^{2} \left( 3003+33915\,{x}^{8}-83980\,{x}^{6}+72930\,{x}^{4}-25740\,{x}^{2} \right) 
$ \\ 
3 & 3 & $3\,{x}^{3}
$ \\ 
3 & 5 & $\frac{1}{2}\sqrt {13}{x}^{3} \left( -9+11\,{x}^{2} \right) 
$ \\ 
3 & 7 & $\frac{1}{8}\sqrt {17}{x}^{3} \left( 99+195\,{x}^{4}-286\,{x}^{2} \right) 
$ \\ 
3 & 9 & $\frac{1}{16}\sqrt {21}{x}^{3} \left( -429+1615\,{x}^{6}-3315\,{x}^{4}+2145\,{x}^{2} \right) 
$ \\ 
4 & 4 & ${x}^{4}\sqrt {11}
$ \\ 
4 & 6 & $\frac{1}{2}\sqrt {15}{x}^{4} \left( -11+13\,{x}^{2} \right) 
$ \\ 
4 & 8 & $\frac{1}{8}\sqrt {19}{x}^{4} \left( 143+255\,{x}^{4}-390\,{x}^{2} \right) 
$ \\ 
4 & 10 & $\frac{1}{16}\sqrt {23}{x}^{4} \left( -715+2261\,{x}^{6}-4845\,{x}^{4}+3315\,{x}^{2} \right) 
$ \\ 
\end{tabular}
\end{ruledtabular}
\end{table}

\subsection{Fourier Transform}
\begin{figure}
\includegraphics[width=8cm,angle=0]{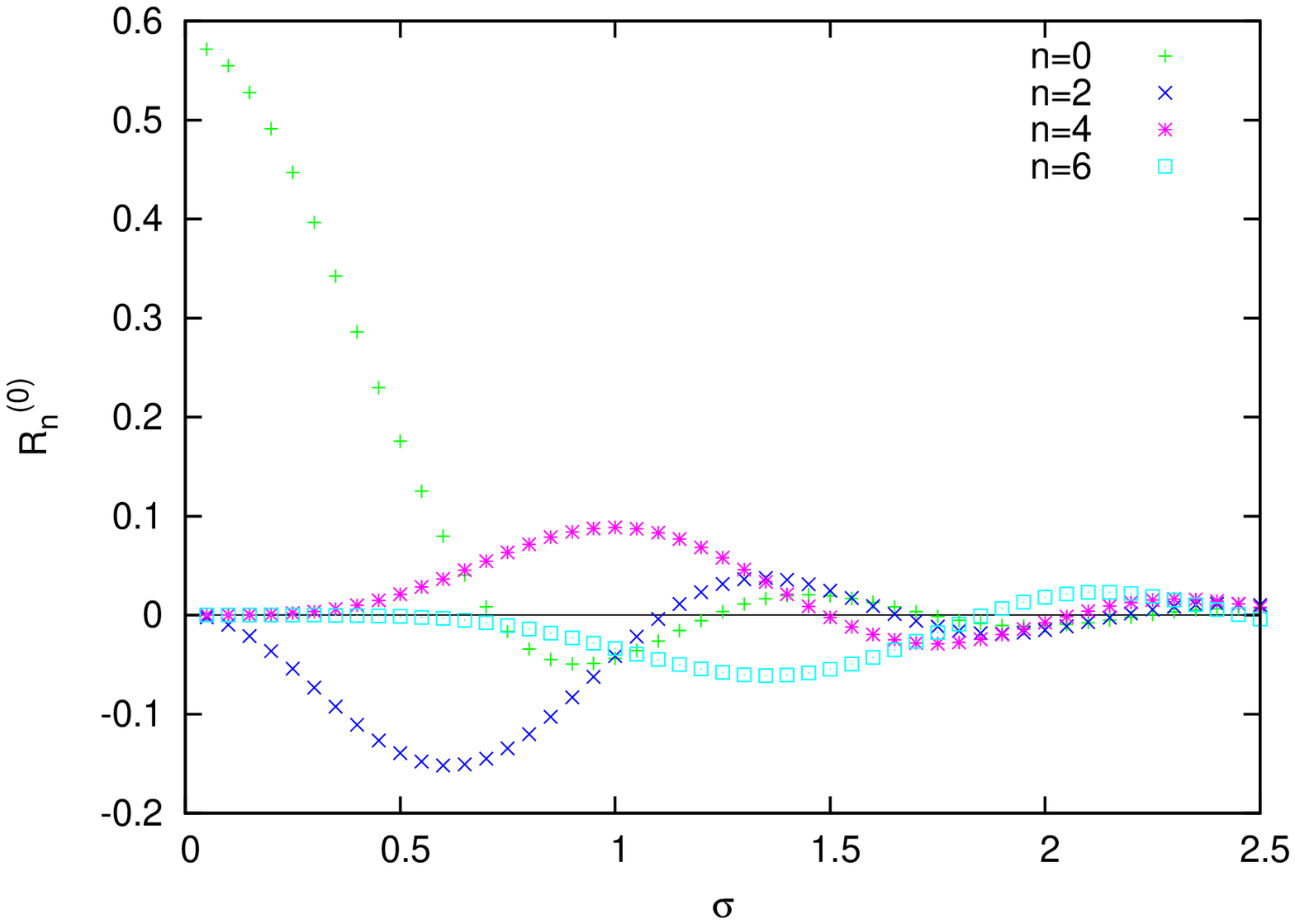}
\includegraphics[width=8cm,angle=0]{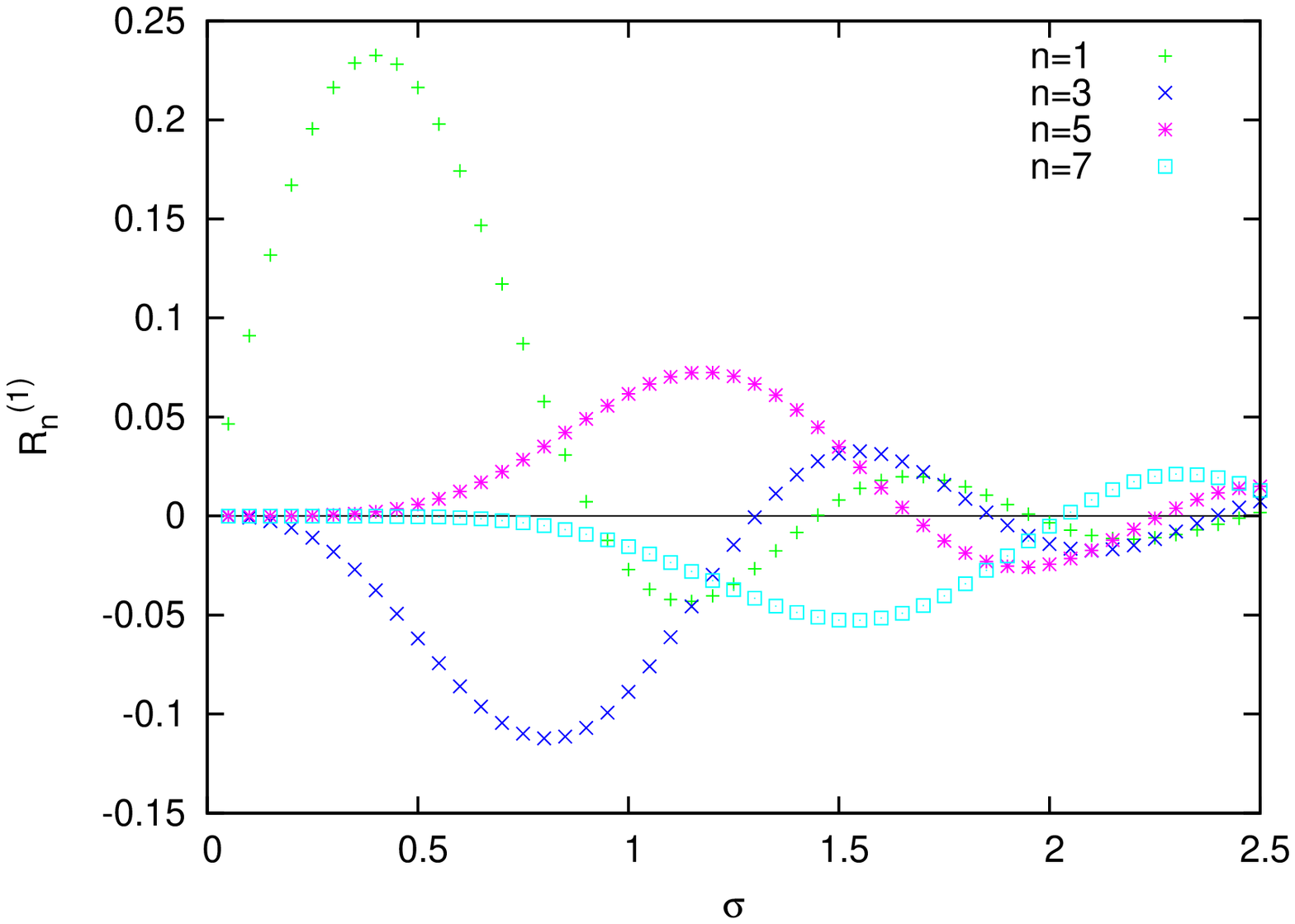}
\includegraphics[width=8cm,angle=0]{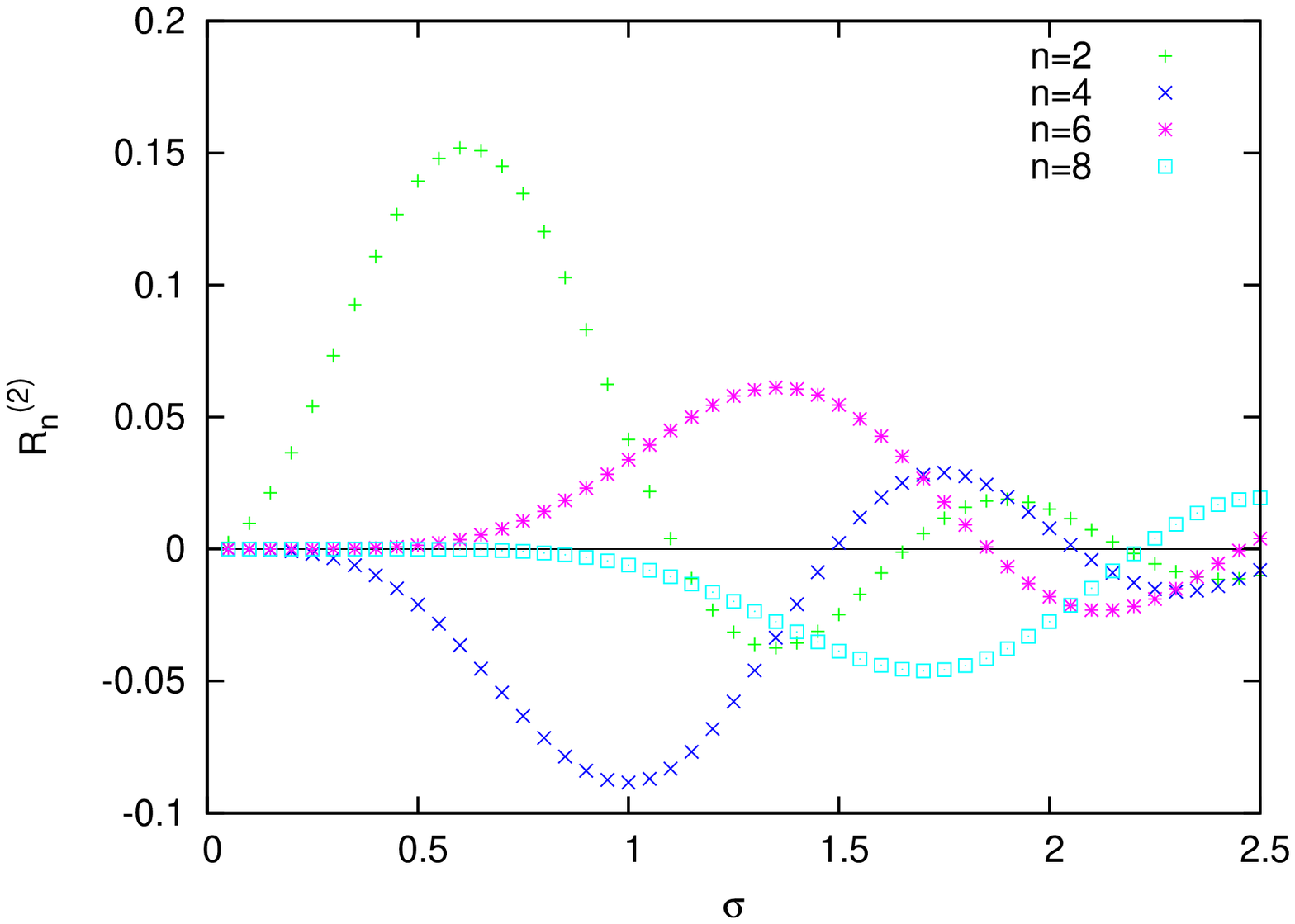}
\includegraphics[width=8cm,angle=0]{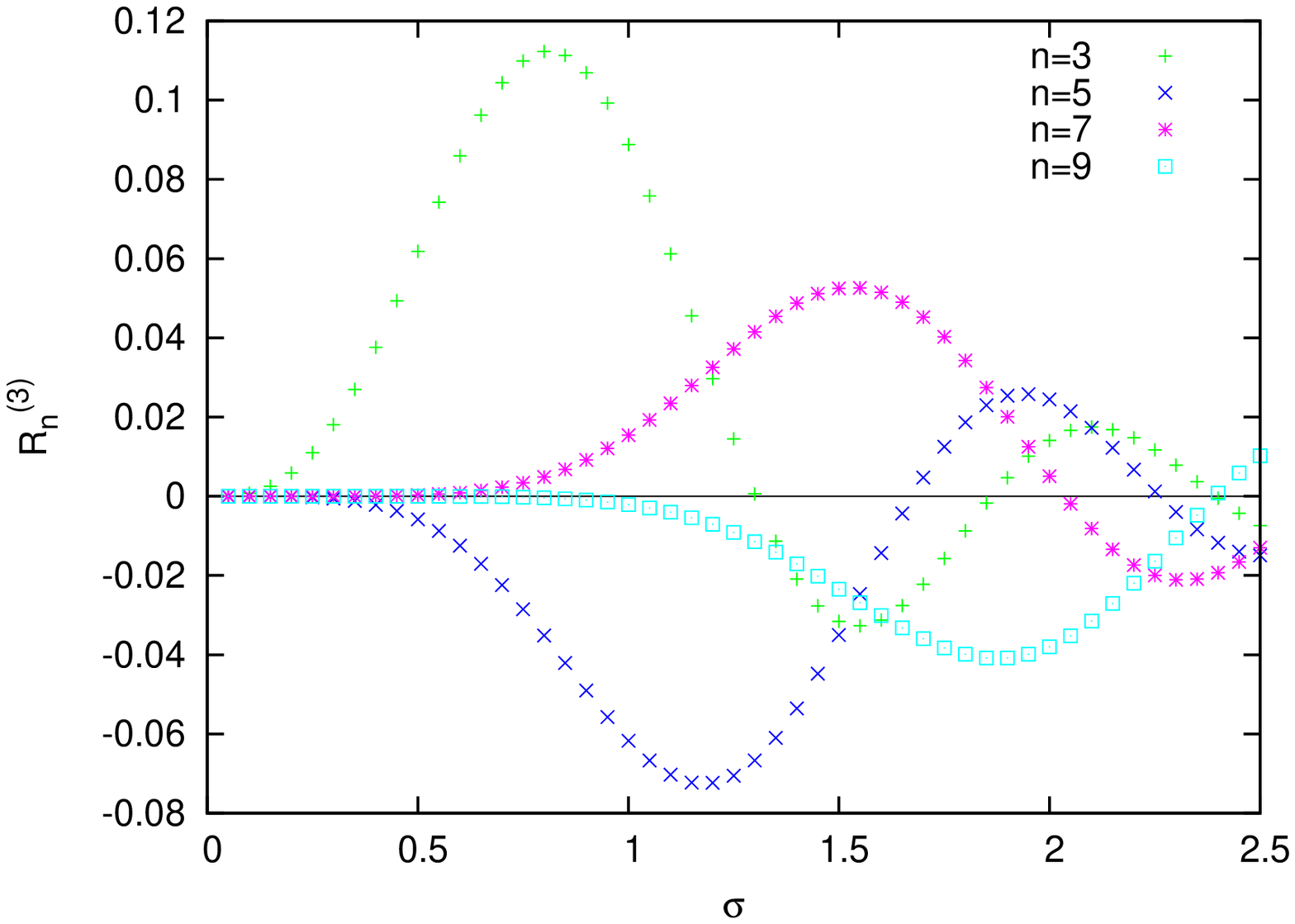}
\caption{The simplest cases of $R_n^{(l)}(\sigma)$, $0\le l\le 3$,
$\alpha\le 3$, as given in (\ref{eq.Rnlsigma}).}
\label{fig.Rsigma}
\end{figure}

The Rayleigh expansion of the plane wave leads to the
Fourier transform in terms of spherical Bessel Functions $j$
\cite[sect.\ 10]{AS},
\begin{equation}
Z_{n,l}^{(m)}(\bm{\sigma})
=
\int d^3x  e^{2\pi i \bm{\sigma}\cdot \bm{x}}R_n^{(l)}(x) Y_l^m(\theta,\phi)
=
4\pi
i^l
Y_l^{m}(\theta_\sigma,\phi_\sigma)
\int_0^1 x^2 dx 
j_l(2\pi \sigma x)
 R_n^{(l)}(x)
.
\label{eq.RFT}
\end{equation}
Repeated use of the partial integration
\cite[11.3.3]{AS}
\begin{equation}
\int x^{\mu+1}\mathcal{C}_{\nu-1}(x)dx = x^{\mu+1}\mathcal{C}_\nu(x)+(\nu-\mu-1)\int x^\mu \mathcal{C}_{\nu}(x)dx
\end{equation}
for any ordinary Bessel Functions $\mathcal{C}$ yields
\begin{equation}
\int_0^1 x^2 dx j_l(2\pi\sigma x)x^{l+2s}
=
\frac{1}{2\pi\sigma}\sum_{k=0}^s (-1)^k\frac{s(s-1)(s-2)\cdots (s-k+1)}{(\pi\sigma)^k}
j_{l+k+1}(2\pi\sigma)
=
\frac{1}{2\pi\sigma}\sum_{k=0}^s \frac{(-s)_k}{(\pi\sigma)^k}
j_{l+k+1}(2\pi\sigma)
.
\end{equation}
This is the building block to transform the radial polynomials of interest, ie,
those constructable from non-negative integers $s$.
Pochhammer's symbol
\cite[6.1.22]{AS}
\begin{equation}
(z)_m\equiv z(z+1)(z+2)\cdots (z+m-1) = \Gamma(z+m)/\Gamma(z);\quad (z)_0\equiv 1,
\end{equation}
has been used to condense the notation for the falling factorials.
Summation over $\alpha$ terms of this type yields
\begin{equation}
\int_0^1 x^2 j_l(2\pi\sigma x)x^l G_\alpha(q,q,x^2) dx
=
\frac{1}{2\pi\sigma}
\frac{\alpha !}{\Gamma(2\alpha+q)}
\sum_{k=0}^{\alpha}
\left( -\frac{1}{\pi\sigma}\right)^k
j_{l+k+1}(2\pi\sigma)
 \Gamma(\alpha+q+k)
{\alpha \choose k} 
.
\end{equation}
Gathering also the constant of normalization in (\ref{eq.Rnorm}) concludes
\begin{eqnarray}
R_n^{(l)}(\sigma)&\equiv &
\int_0^1 x^2 j_l(2\pi\sigma x)R_n^{(l)}(x) dx
=
\frac{\sqrt{2n+3}}{2\pi\sigma}
\sum_{k=0}^{\alpha}
\left( -\frac{1}{\pi\sigma}\right)^k
j_{l+k+1}(2\pi\sigma)
 (\alpha+q)_k
{\alpha \choose k} 
\\
&=&
(-1)^\alpha \sqrt{2n+3}\,\frac{j_{n+1}(2\pi\sigma)}{2\pi\sigma}
,
\label{eq.Rnlsigma}
\end{eqnarray}
orthogonal with
\begin{equation}
\int_0^\infty \sigma^2 R_n^{(l)}(\sigma) R_{n'}^{(l)}(\sigma) d\sigma
=\frac{1}{(4\pi)^2}\delta_{nn'}.
\label{eq.Rsigortho}
\end{equation}
A graphical representation of the cases $l\le 3$, $\alpha\le 3$
is in Figure \ref{fig.Rsigma}.
The leading polynomial order is
$R_n^{(l)}(\sigma)\propto \sigma^n$ as $\sigma\to 0$.
\cite[10.1.2]{AS}.

In summary, (\ref{eq.RFT}) has been factorized as
\begin{equation}
Z_{n,l}^{(m)}(\bm{\sigma}) = 4\pi i^l Y_l^m(\theta_\sigma,\phi_\sigma) R_n^{(l)}(\sigma),
\end{equation}
where Parseval's theorem maintains orthogonality 
in wavenumber space,
\begin{equation}
\int d^3\sigma {Z_{n,l}^{(m)}}^*(\bm{\sigma})Z_{n',l}^{(m')}(\bm{\sigma}) =
\delta_{n,n'}\delta_{m,m'}.
\end{equation}

\section{Mode List} \label{sec.modes}
The dominant modes $K_p^{(l)}$ of (\ref{eq.KofZ}) are listed below,
separated by empty lines, each representing $2l+1$ modes labeled by $m$.
The notation consists of the value $\lambda_{p,l}^2$ followed by a colon,
followed by $l$, followed by another colon and the Zernike expansion.
The Zernike expansion is a sum over floating point numbers $\kappa_{n,l,p}$ multiplied
by $Z_{n,l}^{(m)}$, the latter written as \texttt{Z(n,l)}, optionally
wrapped into a new line. The squares of $\kappa_{n,l,p}$ add to unity in each
line. The phase/sign convention for the eigenvectors is that $\kappa_{n,l,p}$ with
maximum absolute value is positive in each line.

Small-terms contributions with $|\kappa_{n,l,p}|<10^{-6}$ are not printed.
\footnotesize
\begin{verbatim}
0.2723742 :  1:  +0.9935210*Z(1,1) -0.1132657*Z(3,1) +0.0093216*Z(5,1) -0.0001473*Z(7,1) +0.0000128*Z(9,1) 
+0.0000014*Z(11,1)

0.0495069 :  2:  +0.9722349*Z(2,2) -0.2319940*Z(4,2) +0.0305878*Z(6,2) -0.0015909*Z(8,2) +0.0000930*Z(10,2) 
+0.0000023*Z(12,2) +0.0000011*Z(14,2)

0.0495069 :  0:  +0.9722349*Z(2,0) -0.2319940*Z(4,0) +0.0305878*Z(6,0) -0.0015909*Z(8,0) +0.0000930*Z(10,0) 
+0.0000023*Z(12,0) +0.0000011*Z(14,0)

0.0174576 :  3:  +0.9473332*Z(3,3) -0.3153328*Z(5,3) +0.0557030*Z(7,3) -0.0046947*Z(9,3) +0.0003314*Z(11,3) 
-0.0000056*Z(13,3) +0.0000023*Z(15,3)

0.0143736 :  1:  +0.1081177*Z(1,1) +0.9173415*Z(3,1) -0.3757763*Z(5,1) +0.0743888*Z(7,1) -0.0072975*Z(9,1) 
+0.0005484*Z(11,1) -0.0000162*Z(13,1) +0.0000031*Z(15,1)

0.0080675 :  4:  +0.9224733*Z(4,4) -0.3771734*Z(6,4) +0.0818302*Z(8,4) -0.0092993*Z(10,4) +0.0008094*Z(12,4) 
-0.0000345*Z(14,4) +0.0000046*Z(16,4)

0.0059302 :  2:  +0.2144266*Z(2,2) +0.8327378*Z(4,2) -0.4927135*Z(6,2) +0.1320534*Z(8,2) -0.0189879*Z(10,2) 
+0.0019187*Z(12,2) -0.0001183*Z(14,2) +0.0000100*Z(16,2)

0.0059302 :  0:  +0.2144266*Z(2,0) +0.8327378*Z(4,0) -0.4927135*Z(6,0) +0.1320534*Z(8,0) -0.0189879*Z(10,0) 
+0.0019187*Z(12,0) -0.0001183*Z(14,0) +0.0000100*Z(16,0)

0.0043466 :  5:  +0.8987736*Z(5,5) -0.4247285*Z(7,5) +0.1076118*Z(9,5) -0.0151323*Z(11,5) +0.0015784*Z(13,5) 
-0.0000988*Z(15,5) +0.0000094*Z(17,5)

0.0030561 :  3:  +0.2833231*Z(3,3) +0.7478488*Z(5,3) -0.5687499*Z(7,3) +0.1890304*Z(9,3) -0.0349457*Z(11,3) 
+0.0044522*Z(13,3) -0.0003824*Z(15,3) +0.0000310*Z(17,3)

0.0028433 :  1:  +0.0312484*Z(1,1) +0.3323629*Z(3,1) +0.6982757*Z(5,1) -0.5956983*Z(7,1) +0.2106757*Z(9,1) 
-0.0412113*Z(11,1) +0.0054932*Z(13,1) -0.0004989*Z(15,1) +0.0000407*Z(17,1)

0.0025891 :  6:  +0.8765467*Z(6,6) -0.4622201*Z(8,6) +0.1324035*Z(10,6) -0.0219219*Z(12,6) +0.0026626*Z(14,6) 
-0.0002124*Z(16,6) +0.0000189*Z(18,6)

0.0017869 :  4:  +0.3306586*Z(4,4) +0.6685227*Z(6,4) -0.6180700*Z(8,4) +0.2423796*Z(10,4) -0.0539894*Z(12,4) 
+0.0082606*Z(14,4) -0.0008873*Z(16,4) +0.0000803*Z(18,4) -0.0000033*Z(20,4) +0.0000010*Z(22,4)

0.0016562 :  7:  +0.8558181*Z(7,7) -0.4923231*Z(9,7) +0.1559183*Z(11,7) -0.0294312*Z(13,7) +0.0040659*Z(15,7) 
-0.0003881*Z(17,7) +0.0000358*Z(19,7)

0.0015654 :  2:  -0.0803297*Z(2,2) -0.4131467*Z(4,2) -0.5450977*Z(6,2) +0.6585434*Z(8,2) -0.2941911*Z(10,2) 
+0.0731123*Z(12,2) -0.0122479*Z(14,2) +0.0014490*Z(16,2) -0.0001385*Z(18,2) +0.0000078*Z(20,2) 
-0.0000014*Z(22,2)

0.0015654 :  0:  -0.0803297*Z(2,0) -0.4131467*Z(4,0) -0.5450977*Z(6,0) +0.6585434*Z(8,0) -0.2941911*Z(10,0) 
+0.0731123*Z(12,0) -0.0122479*Z(14,0) +0.0014490*Z(16,0) -0.0001385*Z(18,0) +0.0000078*Z(20,0) 
-0.0000014*Z(22,0)

0.0011353 :  5:  -0.3644454*Z(5,5) -0.5960909*Z(7,5) +0.6491312*Z(9,5) -0.2909694*Z(11,5) +0.0750988*Z(13,5) 
-0.0133381*Z(15,5) +0.0016997*Z(17,5) -0.0001754*Z(19,5) +0.0000115*Z(21,5) -0.0000017*Z(23,5)

0.0011176 :  8:  +0.8365082*Z(8,8) -0.5168302*Z(10,8) +0.1780536*Z(12,8) -0.0374644*Z(14,8) 
+0.0057793*Z(16,8) -0.0006360*Z(18,8) +0.0000629*Z(20,8) -0.0000023*Z(22,8) +0.0000010*Z(24,8)

0.0009624 :  3:  -0.1234664*Z(3,3) -0.4534598*Z(5,3) -0.4075425*Z(7,3) +0.6822917*Z(9,3) -0.3674242*Z(11,3) 
+0.1096522*Z(13,3) -0.0219805*Z(15,3) +0.0031596*Z(17,3) -0.0003567*Z(19,3) +0.0000288*Z(21,3) 
-0.0000031*Z(23,3)

0.0009269 :  1:  -0.0132054*Z(1,1) -0.1534280*Z(3,1) -0.4633881*Z(5,1) -0.3595625*Z(7,1) +0.6845747*Z(9,1) 
-0.3857727*Z(11,1) +0.1191289*Z(13,1) -0.0245691*Z(15,1) +0.0036299*Z(17,1) -0.0004190*Z(19,1) 
+0.0000351*Z(21,1) -0.0000037*Z(23,1)

0.0007862 :  9:  +0.8185038*Z(9,9) -0.5369951*Z(11,9) +0.1988032*Z(13,9) -0.0458634*Z(15,9) 
+0.0077854*Z(17,9) -0.0009640*Z(19,9) +0.0001032*Z(21,9) -0.0000060*Z(23,9) +0.0000014*Z(25,9)

0.0007655 :  6:  -0.3891945*Z(6,6) -0.5304668*Z(8,6) +0.6673369*Z(10,6) -0.3345193*Z(12,6) +0.0974584*Z(14,6) 
-0.0196069*Z(16,6) +0.0028703*Z(18,6) -0.0003351*Z(20,6) +0.0000280*Z(22,6) -0.0000032*Z(24,6)

0.0006365 :  4:  -0.1595289*Z(4,4) -0.4702764*Z(6,4) -0.2875787*Z(8,4) +0.6806273*Z(10,4) -0.4291202*Z(12,4) 
+0.1485535*Z(14,4) -0.0344854*Z(16,4) +0.0057898*Z(18,4) -0.0007570*Z(20,4) +0.0000753*Z(22,4) 
-0.0000076*Z(24,4)

0.0005909 :  2:  -0.0387797*Z(2,2) -0.2229036*Z(4,2) -0.4693503*Z(6,2) -0.1796996*Z(8,2) +0.6673417*Z(10,2) 
-0.4675801*Z(12,2) +0.1742073*Z(14,2) -0.0429519*Z(16,2) +0.0076259*Z(18,2) -0.0010468*Z(20,2) 
+0.0001105*Z(22,2) -0.0000110*Z(24,2)

0.0005909 :  0:  -0.0387797*Z(2,0) -0.2229036*Z(4,0) -0.4693503*Z(6,0) -0.1796996*Z(8,0) +0.6673417*Z(10,0) 
-0.4675801*Z(12,0) +0.1742073*Z(14,0) -0.0429519*Z(16,0) +0.0076259*Z(18,0) -0.0010468*Z(20,0) 
+0.0001105*Z(22,0) -0.0000110*Z(24,0)

0.0005719 : 10:  +0.8016857*Z(10,10) -0.5537242*Z(12,10) +0.2182106*Z(14,10) -0.0545025*Z(16,10) 
+0.0100622*Z(18,10) -0.0013779*Z(20,10) +0.0001598*Z(22,10) -0.0000118*Z(24,10) +0.0000020*Z(26,10)

0.0005398 :  7:  -0.4076451*Z(7,7) -0.4711126*Z(9,7) +0.6763013*Z(11,7) -0.3731525*Z(13,7) +0.1204400*Z(15,7) 
-0.0269500*Z(17,7) +0.0044332*Z(19,7) -0.0005783*Z(21,7) +0.0000570*Z(23,7) -0.0000061*Z(25,7)

0.0004436 :  5:  -0.1893914*Z(5,5) -0.4728633*Z(7,5) -0.1841723*Z(9,5) +0.6623861*Z(11,5) -0.4795241*Z(13,5) 
+0.1880973*Z(15,5) -0.0494097*Z(17,5) +0.0094393*Z(19,5) -0.0014015*Z(21,5) +0.0001624*Z(23,5) 
-0.0000171*Z(25,5)

0.0004276 : 11:  +0.7859404*Z(11,11) -0.5676904*Z(13,11) +0.2363442*Z(15,11) -0.0632823*Z(17,11) 
+0.0125856*Z(19,11) -0.0018812*Z(21,11) +0.0002354*Z(23,11) -0.0000206*Z(25,11) +0.0000029*Z(27,11)

0.0004024 :  3:  -0.0651981*Z(3,3) -0.2733637*Z(5,3) -0.4451931*Z(7,3) -0.0326736*Z(9,3) +0.6203023*Z(11,3) 
-0.5285997*Z(13,3) +0.2303104*Z(15,3) -0.0658223*Z(17,3) +0.0135712*Z(19,3) -0.0021567*Z(21,3) 
+0.0002690*Z(23,3) -0.0000292*Z(25,3) +0.0000019*Z(27,3)

0.0003943 :  8:  -0.4215561*Z(8,8) -0.4173790*Z(10,8) +0.6785254*Z(12,8) -0.4071763*Z(14,8) 
+0.1435708*Z(16,8) -0.0352331*Z(18,8) +0.0064074*Z(20,8) -0.0009229*Z(22,8) +0.0001035*Z(24,8) 
-0.0000112*Z(26,8)

0.0003932 :  1:  -0.0068258*Z(1,1) -0.0831481*Z(3,1) -0.2897187*Z(5,1) -0.4325853*Z(7,1) +0.0025750*Z(9,1) 
+0.6065149*Z(11,1) -0.5392619*Z(13,1) +0.2411059*Z(15,1) -0.0703318*Z(17,1) +0.0147685*Z(19,1) 
-0.0023858*Z(21,1) +0.0003027*Z(23,1) -0.0000331*Z(25,1) +0.0000022*Z(27,1)

0.0003271 : 12:  +0.7711645*Z(12,12) -0.5794048*Z(14,12) +0.2532829*Z(16,12) -0.0721245*Z(18,12) 
+0.0153311*Z(20,12) -0.0024758*Z(22,12) +0.0003325*Z(24,12) -0.0000329*Z(26,12) +0.0000043*Z(28,12)

0.0003217 :  6:  -0.2141615*Z(6,6) -0.4666186*Z(8,6) -0.0954608*Z(10,6) +0.6334239*Z(12,6) -0.5195267*Z(14,6) 
+0.2270429*Z(16,6) -0.0663391*Z(18,6) +0.0141531*Z(20,6) -0.0023475*Z(22,6) +0.0003081*Z(24,6) 
-0.0000351*Z(26,6) +0.0000025*Z(28,6)

0.0002963 :  9:  -0.4321073*Z(9,9) -0.3686297*Z(11,9) +0.6757850*Z(13,9) -0.4369714*Z(15,9) 
+0.1665011*Z(17,9) -0.0443179*Z(19,9) +0.0087990*Z(21,9) -0.0013848*Z(23,9) +0.0001726*Z(25,9) 
-0.0000196*Z(27,9) +0.0000011*Z(29,9)

0.0002873 :  4:  +0.0899412*Z(4,4) +0.3087998*Z(6,4) +0.4055501*Z(8,4) -0.0847063*Z(10,4) -0.5563662*Z(12,4) 
+0.5704152*Z(14,4) -0.2846000*Z(16,4) +0.0922604*Z(18,4) -0.0215806*Z(20,4) +0.0038876*Z(22,4) 
-0.0005548*Z(24,4) +0.0000668*Z(26,4) -0.0000059*Z(28,4)

0.0002739 :  2:  +0.0217345*Z(2,2) +0.1326368*Z(4,2) +0.3344713*Z(6,2) +0.3619721*Z(8,2) -0.1549178*Z(10,2) 
-0.5119804*Z(12,2) +0.5869219*Z(14,2) -0.3108572*Z(16,2) +0.1054741*Z(18,2) -0.0256715*Z(20,2) 
+0.0047925*Z(22,2) -0.0007084*Z(24,2) +0.0000877*Z(26,2) -0.0000082*Z(28,2) +0.0000011*Z(30,2)

0.0002739 :  0:  +0.0217345*Z(2,0) +0.1326368*Z(4,0) +0.3344713*Z(6,0) +0.3619721*Z(8,0) -0.1549178*Z(10,0) 
-0.5119804*Z(12,0) +0.5869219*Z(14,0) -0.3108572*Z(16,0) +0.1054741*Z(18,0) -0.0256715*Z(20,0) 
+0.0047925*Z(22,0) -0.0007084*Z(24,0) +0.0000877*Z(26,0) -0.0000082*Z(28,0) +0.0000011*Z(30,0)

0.0002552 : 13:  +0.7572651*Z(13,13) -0.5892625*Z(15,13) +0.2691084*Z(17,13) -0.0809680*Z(19,13) 
+0.0182744*Z(21,13) -0.0031620*Z(23,13) +0.0004537*Z(25,13) -0.0000497*Z(27,13) +0.0000062*Z(29,13)

0.0002407 :  7:  -0.2348116*Z(7,7) -0.4548404*Z(9,7) -0.0194882*Z(11,7) +0.5976955*Z(13,7) -0.5502531*Z(15,7) 
+0.2645281*Z(17,7) -0.0848500*Z(19,7) +0.0199308*Z(21,7) -0.0036443*Z(23,7) +0.0005316*Z(25,7) 
-0.0000659*Z(27,7) +0.0000060*Z(29,7)

0.0002280 : 10:  -0.4401180*Z(10,10) -0.3242846*Z(12,10) +0.6693673*Z(14,10) -0.4629349*Z(16,10) 
+0.1889767*Z(18,10) -0.0540704*Z(20,10) +0.0116042*Z(22,10) -0.0019775*Z(24,10) +0.0002699*Z(26,10) 
-0.0000326*Z(28,10) +0.0000025*Z(30,10)

0.0002127 :  5:  +0.1122926*Z(5,5) +0.3328253*Z(7,5) +0.3585602*Z(9,5) -0.1767793*Z(11,5) -0.4838664*Z(13,5) 
+0.5954557*Z(15,5) -0.3351895*Z(17,5) +0.1212944*Z(19,5) -0.0316439*Z(21,5) +0.0063570*Z(23,5) 
-0.0010169*Z(25,5) +0.0001357*Z(27,5) -0.0000143*Z(29,5) +0.0000018*Z(31,5)

0.0002025 : 14:  +0.7441600*Z(14,14) -0.5975741*Z(16,14) +0.2839011*Z(18,14) -0.0897649*Z(20,14) 
+0.0213924*Z(22,14) -0.0039392*Z(24,14) +0.0006007*Z(26,14) -0.0000717*Z(28,14) +0.0000090*Z(30,14)

0.0001994 :  3:  +0.0387722*Z(3,3) +0.1746705*Z(5,3) +0.3544766*Z(7,3) +0.2795820*Z(9,3) -0.2630382*Z(11,3) 
-0.4046044*Z(13,3) +0.6071129*Z(15,3) -0.3742479*Z(17,3) +0.1447577*Z(19,3) -0.0399704*Z(21,3) 
+0.0084474*Z(23,3) -0.0014183*Z(25,3) +0.0001974*Z(27,3) -0.0000221*Z(29,3) +0.0000026*Z(31,3)

0.0001963 :  1:  +0.0039977*Z(1,1) +0.0501564*Z(3,1) +0.1886747*Z(5,1) +0.3560673*Z(7,1) +0.2581983*Z(9,1) 
-0.2814148*Z(11,1) -0.3836754*Z(13,1) +0.6083843*Z(15,1) -0.3837324*Z(17,1) +0.1508905*Z(19,1) 
-0.0422465*Z(21,1) +0.0090396*Z(23,1) -0.0015357*Z(25,1) +0.0002159*Z(27,1) -0.0000245*Z(29,1) 
+0.0000029*Z(31,1)

0.0001848 :  8:  +0.2521278*Z(8,8) +0.4396161*Z(10,8) -0.0455476*Z(12,8) -0.5579080*Z(14,8) 
+0.5728641*Z(16,8) -0.2999791*Z(18,8) +0.1045395*Z(20,8) -0.0267374*Z(22,8) +0.0053315*Z(24,8) 
-0.0008530*Z(26,8) +0.0001150*Z(28,8) -0.0000121*Z(30,8) +0.0000016*Z(32,8)

0.0001789 : 11:  -0.4461735*Z(11,11) -0.2838305*Z(13,11) +0.6602212*Z(15,11) -0.4854523*Z(17,11) 
+0.2108168*Z(19,11) -0.0643653*Z(21,11) +0.0148112*Z(23,11) -0.0027123*Z(25,11) +0.0004011*Z(27,11) 
-0.0000515*Z(29,11) +0.0000047*Z(31,11)

0.0001630 : 15:  +0.7317763*Z(15,15) -0.6045873*Z(17,15) +0.2977377*Z(19,15) -0.0984783*Z(21,15) 
+0.0246634*Z(23,15) -0.0048055*Z(25,15) +0.0007755*Z(27,15) -0.0000998*Z(29,15) +0.0000127*Z(31,15)

0.0001621 :  6:  +0.1321892*Z(6,6) +0.3482620*Z(8,6) +0.3088642*Z(10,6) -0.2477773*Z(12,6) -0.4081762*Z(14,6) 
+0.6063296*Z(16,6) -0.3809198*Z(18,6) +0.1519987*Z(20,6) -0.0436556*Z(22,6) +0.0096556*Z(24,6) 
-0.0017060*Z(26,6) +0.0002504*Z(28,6) -0.0000300*Z(30,6) +0.0000036*Z(32,6)

0.0001501 :  4:  +0.0559858*Z(4,4) +0.2088990*Z(6,4) +0.3571657*Z(8,4) +0.1959894*Z(10,4) -0.3351164*Z(12,4) 
-0.2950686*Z(14,4) +0.6047083*Z(16,4) -0.4289290*Z(18,4) +0.1863726*Z(20,4) -0.0574666*Z(22,4) 
+0.0135356*Z(24,4) -0.0025371*Z(26,4) +0.0003927*Z(28,4) -0.0000500*Z(30,4) +0.0000060*Z(32,4)
\end{verbatim}
\normalsize

\bibliographystyle{apsrmp}
\bibliography{all}

\end{document}